\documentclass[10pt,twocolumn,letterpaper]{article}

\usepackage[pagenumbers]{cvpr} % To force page numbers, e.g. for an arXiv version

\usepackage{booktabs}  % for \toprule, \midrule, \bottomrule
\usepackage{multirow}  % for multi-row cells
\usepackage{makecell}  % for \makecell
\usepackage[table]{xcolor}  % for \rowcolor and gray backgrounds
\usepackage{xcolor} % add this in the preamble
\definecolor{cvprblue}{rgb}{0.21,0.49,0.74}
\usepackage[pagebackref,breaklinks,colorlinks,allcolors=cvprblue]{hyperref}

\def\paperID{27197} % *** Enter the Paper ID here
\def\confName{CVPR}
\def\confYear{2026}

\title{FedPoisonTTP: A Threat Model and Poisoning Attack for Federated Test-Time Personalization}

\author{
Md Akil Raihan Iftee$^{1,}$\thanks{Corresponding author: \href{mailto:iftee1807002@gmail.com}{iftee1807002@gmail.com}}, 
Syed Md.~Ahnaf Hasan$^{1}$, 
Amin Ahsan Ali$^{1}$, 
A K M Mahbubur Rahman$^{1}$,\\
Sajib Mistry$^{2}$, 
Aneesh Krishna$^{2}$\\
\\
$^1$Center for Computational \& Data Sciences (CCDS), Independent University, Bangladesh\\
$^2$Curtin University, Australia\\
}

\begin{document}
\maketitle
\begin{abstract}
Test-time personalization in federated learning enables models at clients to adjust online to local domain shifts, enhancing robustness and personalization in deployment. Yet, existing federated learning work largely overlooks the security risks that arise when local adaptation occurs at test time. Heterogeneous domain arrivals, diverse adaptation algorithms, and limited cross-client visibility create vulnerabilities where compromised participants can craft poisoned inputs and submit adversarial updates that undermine both global and per-client performance. To address this threat, we introduce \textbf{FedPoisonTTP}, a realistic grey-box attack framework that explores test-time data poisoning in the federated adaptation setting. FedPoisonTTP distills a surrogate model from adversarial queries, synthesizes in-distribution poisons using feature-consistency, and optimizes attack objectives to generate high-entropy or class-confident poisons that evade common adaptation filters. These poisons are injected during local adaptation and spread through collaborative updates, leading to broad degradation. Extensive experiments on corrupted vision benchmarks show that compromised participants can substantially diminish overall test-time performance.

\end{abstract}
\section{Introduction}
 Smartphones, browsers, and a growing array of internet-connected sensors and Web-of-Things (WoT) gadgets generate a large share of global network traffic and continuously produce diverse, user-generated content and telemetry. This stream of heterogeneous data is leveraged by machine-learning systems that power many web services, personalized recommendation \cite{ying2018graph, covington2016deep}, social feeds \cite{yu2021self}, and content moderation \cite{pavlopoulos2017deep, badjatiya2017deep} among them. Because these models rely on information produced or held on users’ devices, they create important privacy and data-control challenges for both users and service operators \cite{zhao2025federation, huang2021evaluating, abadi2016deep}. 
 In this regard, Federated Learning (FL) has emerged as a practical paradigm for training machine learning models across decentralized clients while preserving data locality and user privacy \cite{mcmahan2017communication, bonawitz2021federated, tan2022towards}. Additionally, in many real-world deployments the data seen by each client is nonstationary: clients encounter domain shifts over time due to changing sensors, environments, user behaviors, or acquisition conditions. Hence, Test-Time Adaptation (TTA) is necessary to adapt the models in real time to the distribution shifts of the incoming data \cite{wang2021tent}. To maintain performance in the presence of such shifts in collaborative learning, recent work \cite{rajib2025fedctta} has explored \emph{federated test-time adaptation} (FTTA), a hybrid workflow in which a shared global model is periodically aggregated on the server while each client performs lightweight, unsupervised adaptation on its own incoming unlabeled stream at test time. By combining federation for shared representation learning with per-client test-time adaptation (TTA) routines (for example, batch-normalization (BN) adaptation, entropy minimization, or self-supervised consistency updates) \cite{wang2021tent, kang2024membn, ma2024improved}, FTTA provides a practical path to personalized, robust inference without accessing or transmitting raw client data.

Despite these practical advantages, both federated learning and test-time adaptation are vulnerable to adversarial attacks \cite{kumar2023impact, bhagoji2019analyzing, wu2023uncovering, rifatadversarial}. Federated systems are vulnerable to classical and model-poisoning attacks: compromised clients can send malicious updates to skew the global model, leak information, or induce targeted misclassification \cite{bagdasaryan2020backdoor}. Many defense strategies, like robust aggregation, clipping, differential privacy, etc., reduce but do not fully eliminate these risks, and they often trade off utility \cite{nair2023robust, li2023enhancing, sun2020ldp}. Meanwhile, TTA algorithms rely on online unsupervised objectives such as entropy minimization or dynamic BN updates and therefore, can be manipulated by carefully crafted inputs that steer the adaptation process toward harmful local minima \cite{croce2022evaluating}. Prior work in the centralized setup has demonstrated that an adversary with strong access (white-box, or with knowledge of a held-out benign validation set) can craft per-example perturbations that derail TTA \cite{cong2024test, wu2023uncovering}. However, these threat models do not capture a realistic federated deployment: an adversarial client in FL typically cannot observe other clients' data or per-client updates, interacts with the server only via broadcast/aggregation, and participates intermittently with a limited budget for poisoning.

Bridging these two literatures, we study how a constrained adversary operating as one (or a few) malicious participants in an FTTA system can mount a highly effective and stealthy poisoning attack. The key challenge is \emph{realism}: the attacker must succeed despite only seeing the global model broadcasts, lacking access to other clients' local data or updates, and being limited in how many and which rounds they can influence. Drawing inspiration from recent single-node online poisoning approaches \cite{rifatadversarial, wu2023uncovering}, we develop a practical federated attack that rests on three core ideas. First, the adversary builds and continually refines a lightweight \emph{surrogate aggregator} that predicts the post-aggregation model dynamics by combining history-based extrapolation of past global updates with posterior distillation on locally observed responses. Second, the adversary crafts \emph{in-distribution} poisoned examples whose intermediate feature statistics are constrained to match those of a benign local pool via a feature-distribution regularizer; this encourages the harmful perturbations to transfer across heterogeneous client domains and to evade simple detection heuristics. Third, the attack optimizes TTA-aware objectives that explicitly target common adaptation rules deployed by clients (for example, shifting BN channel statistics to mislead BN-adaptive TTA, or shaping entropy/confidence in a class-balanced way to subvert entropy-minimizing updates). All components are designed to preserve stealth: per-example perturbation bounds, class balance, and upload-norm clipping are enforced so that the poisoned updates resemble ordinary client updates and evade naïve aggregation defenses.

This paper makes the following contributions:

\begin{itemize}
\item We formalize the problem of poisoning in federated test-time adaptation and introduce a realistic threat model where the adversary is a limited federated client that cannot access other clients' data or local models and only observes global model broadcasts. This threat model fills an important gap between single-node TTA attacks and conventional FL poisoning scenarios.
\item We propose a practical attack framework that combines history-based surrogate aggregation, posterior distillation, and a feature-distribution regularizer to craft in-distribution poisoned examples that transfer across heterogeneous client domains. The attack is TTA-aware: it explicitly targets BN-based adaptation and entropy-driven updates that are commonly used in FTTA.
\item We provide an algorithmic implementation that respects strict stealth constraints (perturbation bounds, class-balancing, and update-norm clipping) and is compatible with standard FL pipelines (partial participation, FedAvg-style aggregation, and diverse per-client TTA strategies).
\item Through a comprehensive empirical evaluation on federated domain-shift benchmarks and a variety of client TTA methods, we demonstrate that our attack produces substantial and persistent degradation across many honest clients while remaining difficult to detect by basic defensive measures. We further analyze key design choices to understand which components most significantly contribute to the attack’s effectiveness.
\end{itemize}
\section{Related Work}
\subsection{Federated learning and personalization}
Federated learning (FL) enables collaborative model training across many clients without centralizing raw data. This promotes privacy among the clients as they do not need to share their personal data. Federated Learning was proposed in the work of \cite{mcmahan2017communication} in the form of FedAvg algorithm, which uses practical iterative averaging for decentralized learning. Subsequent works have further examined the challenges of handling heterogeneous updates \cite{li2020federated}, developed mechanisms for client-drift correction \cite{karimireddy2020scaffold}, proposed methods for improving representation alignment \cite{li2021model}, and introduced strategies for objective normalization \cite{wang2020tackling}.

Personalized federated learning (PFL) has emerged to improve per-client performance under heterogeneity. Approaches include meta-learning / Per-FedAvg-style schemes that learn initializations well-suited for local finetuning \cite{fallah2020personalized}, Moreau-envelope regularization (pFedMe) for decoupling personalized objectives \cite{dinh2020pfedme}, and many hybrid strategies that trade global generalization for per-client specialization. Practical deployments also require robustness to malicious or unreliable participants; a sizeable literature on robust aggregation and Byzantine-resilient methods has therefore developed (e.g., Krum, geometric-median/RFA, center-clipping and momentum-based defenses), as well as defenses aimed specifically at Sybil or backdoor-style threats such as FoolsGold \cite{blanchard2017machine, pillutla2019robust, karimireddy2021learning, fung2020limitations}. Despite these defenses, powerful models, and data-poisoning attacks (including model replacement and stealthy backdoors) remain effective in many realistic federated settings \cite{bagdasaryan2020backdoor, baruch2019little}.

\subsection{Test-time adaptation}
Test-time adaptation (TTA) techniques adapt pretrained models at inference to mitigate distribution shift without access to source labels. Early practical methods showed that updating normalization statistics or simple self-supervised objectives at test time yields substantial improvements \cite{li2016revisiting, sun2020test}. \emph{Tent} (entropy minimization with BN-affine updates) is a widely-adopted TTA approach that optimizes prediction entropy and BN statistics online \cite{wang2021tent}; related source-free and pseudo-labeling approaches (e.g., SHOT) adopt entropy / information-maximization objectives and feature clustering for target adaptation \cite{liang2020shot}. TTA methods can be very lightweight and effective across corrupted/shifted benchmarks, which is why they are attractive for per-client personalization in federated deployments. However, their online, label-free nature also changes the attack surface.

\subsection{Adversarial attacks on test-time adaptation}
TTA’s runtime updates open new avenues for adversaries. Classic data and model-poisoning literature (e.g., targeted clean-label poisoning, poisoning SVMs) established that training-time data manipulation can cause severe downstream errors \cite{biggio2012poisoning, shafahi2018poison}. More recently, a host of works have analyzed and demonstrated test-time poisoning—where an adversary injects malicious inputs during deployment to steer online adaptation. Notable attacks include the Distribution Invading Attack (DIA), which shows how inserting a small fraction of malicious examples into test batches can hijack BN-based and entropy-based TTA methods \cite{wu2023uncovering}, and TePA, which performs surrogate-model-based test-time poisoning and demonstrates transferability across TTA variants \cite{cong2023tepa}. Empirical studies show that only a few poisoned test samples can substantially degrade TTA performance on standard corrupted benchmarks \cite{cong2023tepa, wu2023uncovering}.

\section{Problem Setup and Threat Model}
\subsection{Federated Test-Time Adaptation}
We consider a federated learning (FL) system with $N$ clients coordinated by a central server. At round $t$, the server maintains a global model $\theta^t$, which is broadcast to a subset of participating clients $S_t \subseteq \{1,\ldots,N\}$. Each client $i \in S_t$ receives $\theta^t$ and adapts it to its local unlabeled stream $\mathcal{D}_{i,t}$ using a test-time adaptation (TTA) algorithm $T_i$, such as entropy minimization \cite{wang2021tent}, batch-normalization (BN) adaptation, or self-training methods \cite{liang2020shot}. After adaptation, the client returns either the updated model parameters $\theta^t_i$ or their difference $\Delta \theta^t_i$ to the server, which aggregates the received updates (e.g., via FedAvg) to produce the new global model:
\begin{equation}
    \theta^{t+1} = \theta^t + \eta \sum_{i \in S_t} \frac{n_i}{\sum_{j \in S_t} n_j} \Delta \theta^t_i,
\end{equation}
where $n_i$ denotes the effective sample size for client $i$.  
This hybrid procedure of federated aggregation coupled with per-client TTA which is referred to as \emph{Federated Test-Time Adaptation (FTTA)} and is designed to provide both personalization and robustness to domain shifts.

\subsection{Adversarial Federated Client}
We study an adversary controlling a single client $a \in {1,\ldots,N}$, called the \emph{adversarial client}. Unlike prior poisoning works assuming white-box access to all updates or centralized validation sets, our attacker operates under a \emph{realistic, constrained} setting:
\begin{itemize}
    \item \textbf{Limited visibility:} the attacker observes only the broadcast global model $\theta^t$ at each round. It has no access to other clients’ local data, their local updates, or the details of their adaptation trajectories.
    \item \textbf{Local control:} the attacker controls its own local data stream $\mathcal{D}_{a,t}$ and local training process. It can inject poisoned examples $\mathcal{D}_{a,t}^p \subset \mathcal{D}_{a,t}$ or perturb existing inputs, but it must also handle its remaining benign samples.
    \item \textbf{Participation constraints:} the attacker may not participate in every round; its participation is stochastic and governed by the server’s sampling procedure. This models realistic partial participation in FL.
    \item \textbf{Stealth requirements:} the attacker’s local updates must remain within norm bounds to avoid detection. Perturbations to inputs are constrained by $\ell_\infty$ or $\ell_2$ budgets, and label distributions must remain balanced to evade anomaly detection.
\end{itemize}

\begin{figure*}[!t]
    \centering
    \includegraphics[width=0.9\linewidth, keepaspectratio]{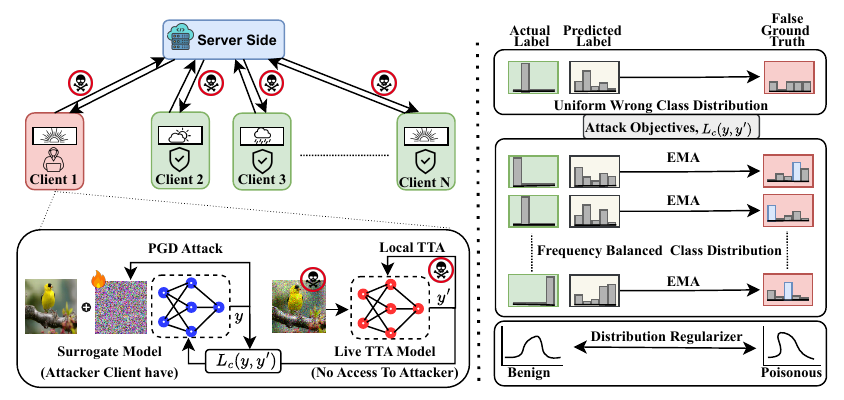}
    
    \caption{Overview of the FedPoisonTTP threat model. A subset of clients is compromised and injects poisoned inputs during local test-time adaptation. The attacker uses a surrogate model to craft in-distribution poisons via PGD-based optimization, which are then fed into the live TTA model, causing corrupted updates that propagate through the federated aggregation process and degrade the performance of honest clients.}
    \label{fig:proposed_approach}
    \vspace{-0.3cm}
\end{figure*}

\subsection{Attack Objective}
The goal of the attacker is to induce a noticeable degradation of performance on the post-aggregation models of benign clients, measured on their respective unseen local data streams. Formally, let $B_{i,t}$ denote a benign batch from client $i$ at round $t$. After aggregation and local adaptation, the benign client evaluates using $h(x;\theta^{t+1}_i)$, where $\theta^{t+1}i = T_i(\theta^{t+1}, B{i,t})$. The adversary seeks to maximize the expected error across all benign clients:
\begin{equation}
    \max_{\mathcal{D}_{a,t}^p} \quad 
    \mathbb{E}_{i \neq a}\Big[\mathcal{L}\big(h(x;\theta^{t+1}_i), y\big)\Big],
\end{equation}
subject to the attacker’s constraints on participation, perturbation budgets, and update norms.  

Since the attacker cannot access $B_{i,t}$ or $\theta^t_i$, it must rely on surrogates: a local clean pool $\mathcal{B}_{ab}$ that approximates benign distributions and a \emph{surrogate aggregator} $\hat{\theta}^{t+1}$ constructed from historical global updates.  

\subsection{Realistic Federated Test-Time Poisoning}
To summarize, our proposed attack protocol satisfies the following realism criteria:
\begin{enumerate}
    \item \textbf{Federated grey-box:} the attacker only observes the global model and its own local training process.
    \item \textbf{No benign access:} the attacker cannot use other clients’ data or updates in its optimization.
    \item \textbf{Online and round-limited:} poisoning occurs during the attacker’s normal participation rounds; no retrospective modification is allowed.
    \item \textbf{Stealthy influence:} the poisoned updates are carefully norm-constrained, class-balanced, and maintained in-distribution, so as to minimize the risk of detection.
\end{enumerate}

We refer to this setting as \emph{Realistic Federated Test-Time Poisoning (R-FTTP)}. In the following section, we outline and develop a set of attack strategies designed to exploit the underlying TTA dynamics within this threat model.

\section{Methods}
In this section we develop a practical attack framework for Realistic Federated Test-Time Poisoning (R-FTTP). Our approach extends single-node RTTDP ideas \cite{su2024adversarial} to the federated setting by (a) building a lightweight surrogate aggregator that predicts the post-aggregation model available to honest clients, (b) crafting \emph{in-distribution} poisoned examples whose intermediate-feature statistics are constrained to match a benign local pool, and (c) optimizing \emph{TTA-aware} objectives that directly target common client adaptation mechanisms (e.g., BN adaptation, entropy minimization). We first give an overview, then present the surrogate aggregator, the feature-regularized poisoning formulation, the TTA-aware objectives, and finally an end-to-end optimization algorithm with practical constraints.

\subsection{Overview}
At each federated round $t$ the server broadcasts the global model $\theta^t$ to a subset of clients $S_t$. Each client $i\in S_t$ performs local test-time adaptation $T_i$ on its incoming unlabeled stream $\mathcal{D}_{i,t}$ and computes a local update $\Delta\theta^t_i$ that is returned to the server for aggregation. Our adversarial client $a$ seeks to craft a poisoned local dataset $\mathcal{D}_{a,t}^p$ (a small fraction of its training stream) so that after aggregation and subsequent local TTA on other clients, honest clients suffer substantial degradations. Because the attacker cannot access other clients' data or raw updates, it relies on (i) a \emph{surrogate aggregator} $\hat{\theta}^{t+1}$ that approximates the true post-aggregation model, and (ii) a local benign pool $\mathcal{B}_{ab}$ to estimate cross-client transfer. The attack optimizes the poisoned inputs through projected gradient steps in order to maximize a surrogate outer loss evaluated on $\hat{\theta}^{t+1}$, while simultaneously enforcing a feature‑distribution regularizer to preserve transferability and maintain stealth.

\subsection{Surrogate Aggregator}
The attacker cannot observe other clients' contributions to aggregation. We therefore define a surrogate aggregator that combines (A) a history-based estimate of other-clients' average update and (B) a posterior-distillation step that refines the surrogate using the broadcasted global model.

\paragraph{History-based aggregation estimate.} The attacker stores the last $k$ global models $\{\theta^{t-k+1},\dots,\theta^t\}$ and forms a historical average update
\begin{equation}
    \widehat{\Delta\theta}_{\mathrm{hist}} \;=\; \frac{1}{k}\sum_{j=t-k+1}^{t} (\theta^{j} - \theta^{j-1}) .
\end{equation}
Assuming other clients' average update does not change rapidly, we estimate the other-clients' contribution at round $t$ by
\begin{equation}
    \widehat{\Delta\theta}_{-a} \;\approx\; \widehat{\Delta\theta}_{\mathrm{hist}} - \widehat{\Delta\theta}_{a,\mathrm{est}},
\end{equation}
where $\widehat{\Delta\theta}_{a,\mathrm{est}}$ is the attacker's estimate of its own historical contribution (maintained by the attacker).

\paragraph{Candidate post-aggregation surrogate.} Given a candidate local attacker update $\Delta\theta_a$ (obtained by training on a candidate poisoned set $\mathcal{D}_{a,t}^p$), the attacker constructs a surrogate post-aggregation model
\begin{equation}
    \hat{\theta}^{t+1}(\Delta\theta_a) \;=\; \theta^t + \eta\big( \widehat{\Delta\theta}_{-a} + \Delta\theta_a \big),
    \label{eq:hat_theta}
\end{equation}
where $\eta$ is the aggregation scaling (server learning rate or weighting factor). Equation~\eqref{eq:hat_theta} is intentionally simple and inexpensive to compute; in practice the attacker refines $\widehat{\Delta\theta}_{-a}$ online by comparing its surrogate predictions to the actual broadcast $\theta^{t+1}$ once available (posterior distillation), i.e., minimizing a distillation loss on the attacker’s benign pool $\mathcal{B}_{ab}$. Concretely, after receiving $\theta^{t+1}$ the attacker may update an internal state $s$ by minimizing
% \begin{equation}
%     \mathcal{L}_{\mathrm{distill}}(s) \;=\; \frac{1}{|\mathcal{B}_{ab}|}\sum_{x\in\mathcal{B}_{ab}} \mathrm{KL}\big( h(x;\theta^{t+1}) \,\|\, h(x;\hat{\theta}^{t+1}(s)) \big),
% \end{equation}

\begin{equation}
    \resizebox{0.9\linewidth}{!}{$\displaystyle
    \mathcal{L}_{\mathrm{distill}}(s) \;=\; \frac{1}{|\mathcal{B}_{ab}|}\sum_{x\in\mathcal{B}_{ab}} \mathrm{KL}\big( h(x;\theta^{t+1}) \,\|\, h(x;\hat{\theta}^{t+1}(s)) \big)
    $},
\end{equation}
where $h(\cdot;\theta)$ are model posteriors and $s$ parameterizes the surrogate dynamics (e.g., corrections to $\widehat{\Delta\theta}_{-a}$). This posterior distillation improves future surrogate quality while respecting the attacker's limited visibility.

\subsection{In-distribution Poisoning with Feature Regularizer}
A key challenge is ensuring that poisoned examples generated on the attacker’s local distribution transfer to unseen client domains. To address this, we constrain poisoning to remain \emph{in‑distribution} by matching the intermediate feature statistics of the poisoned samples to those of a small local benign pool $\mathcal{B}_{ab}$, thereby maintaining alignment with the underlying data distribution.

\paragraph{Features and sample statistics.} Let $f^l(x;\theta)$ denote the activation (flattened over spatial dimensions) at layer $l$ for input $x$. For a sample $x$ we compute its feature mean and (diagonal) variance at layer $l$:
% \begin{equation}
%     \mu^l(x) \;=\; \frac{1}{M_l}\sum_{m=1}^{M_l} f^l_m(x), \qquad
%     \sigma^l(x)^2 \;=\; \frac{1}{M_l}\sum_{m=1}^{M_l} \big(f^l_m(x)-\mu^l(x)\big)^2,
% \end{equation}
\begin{equation}
\begin{split}
    \mu^l(x) &\;=\; \frac{1}{M_l}\sum_{m=1}^{M_l} f^l_m(x), \\
    \sigma^l(x)^2 &\;=\; \frac{1}{M_l}\sum_{m=1}^{M_l} \big(f^l_m(x)-\mu^l(x)\big)^2,
\end{split}
\end{equation}

where $M_l$ is the number of flattened feature dimensions at layer $l$. For a set $\mathcal{S}$ we denote the empirical mean and variance by $\mu^l(\mathcal{S})$ and $\sigma^l(\mathcal{S})^2$.

\paragraph{Feature-distribution regularizer.} For a candidate poisoned set $\mathcal{D}_{a,t}^p$ we define a layer-averaged moment-matching regularizer:
\begin{equation}
\begin{aligned}
    \mathcal{L}_{\mathrm{reg}}(\mathcal{D}_{a,t}^p,\mathcal{B}_{ab}) 
    \;=\; \frac{1}{L}\sum_{l=1}^{L} \Big(& \| \mu^l(\mathcal{D}_{a,t}^p) - \mu^l(\mathcal{B}_{ab}) \|_2^2 \\
    &+ \beta\| \sigma^l(\mathcal{D}_{a,t}^p) - \sigma^l(\mathcal{B}_{ab}) \|_2^2 \Big),
\end{aligned}
\label{eq:Lreg}
\end{equation}

where $L$ is the number of selected layers and $\beta>0$ balances mean vs variance matching. Using diagonal moments keeps computation inexpensive and stable; in practice we select a small set of early- and mid-level layers (e.g., conv1, conv3, a penultimate block) that capture domain characteristics relevant to many clients.

\subsection{TTA-aware Attack Objectives}
We target two complementary TTA pathways commonly used by clients: entropy-minimization-style updates (e.g., TENT) and BN/statistics adaptation. Each objective is optimized while keeping $\mathcal{L}_{\mathrm{reg}}$ small to ensure transferability.

\paragraph{BN-shift objective.} BN-based TTA methods adapt to incoming batch statistics; by carefully shifting channel-wise means/variances the attacker can bias BN-affine parameters or running-statistics in the aggregated model. Let $g_{\mathrm{BN}}(\theta,x)$ denote the vector of channel means/variances at selected early layers for input $x$. The BN-shift objective encourages a targeted shift $s$ in channel statistics:
\begin{equation}
    \mathcal{L}_{\mathrm{BN}}(\mathcal{D}_{a,t}^p;\theta) \;=\; - \frac{1}{|\mathcal{D}_{a,t}^p|}\sum_{x\in\mathcal{D}_{a,t}^p} \langle g_{\mathrm{BN}}(\theta,x) , s \rangle,
    \label{eq:bn_obj}
\end{equation}
where the sign is chosen so that local updates foster a persistent shift in the aggregated model’s BN-related parameters. The target shift $s$ may be estimated from the attacker’s surrogate runs (for instance, directions known to degrade validation accuracy under simulated TTA).

%%%%% BOTH COTTA & TENT TABLE %%%%%
\begin{table*}[!t]
\centering
\caption{Comparison of federated learning (FL) methods with different test-time adaptation (TTA) techniques on CIFAR-10-C and CIFAR-100-C benchmark datasets with poison ratio, $\alpha$ = 0.5. The total number of clients is considered to be N = 10, out of which 5 (M = 5) are considered to be malicious. The poison ratio indicates the ratio of samples that are corrupted in the batches of the malicious clients.}
\label{tab:federated_tta}
\resizebox{0.6\textwidth}{!}{%
\begin{tabular}{llccc@{\hskip 6pt}ccc}
\toprule
\multirow{3}{*}[-0.7em]{\makecell{\textbf{Fed} \\ \textbf{Method}}} & \multirow{3}{*}[-0.7em]{\makecell{\textbf{TTA} \\ \textbf{Method}}}
& \multicolumn{3}{c}{\textbf{CIFAR-10-C}}
& \multicolumn{3}{c}{\textbf{CIFAR-100-C}} \\
\cmidrule(lr){3-5} \cmidrule{6-8}
& & \multirow{2}{*}{\makecell{Clean}} & \multicolumn{2}{c}{Attack Scenario}
  & \multirow{2}{*}{\makecell{Clean}} & \multicolumn{2}{c}{Attack Scenario} \\
\cmidrule(lr){4-5} \cmidrule{7-8}
& & & \makecell{Benign \\ Clients} & \makecell{Adversarial \\ Clients}
  & & \makecell{Benign \\ Clients} & \makecell{Adversarial \\ Clients} \\
\midrule
Any & Source & 56.51 & 56.66 & 56.05 & 54.28 & 53.24 & 52.86 \\
\midrule
\multirow{2}{*}{None} 
    & Tent   & 81.08 & 80.97  & 74.32 & 68.59 & 67.18 & 52.92 \\
    & CoTTA  & 81.78 & 78.45 & 69.21 & 66.24 & 61.73 & 48.15 \\
\midrule
\multirow{2}{*}{FedAvg} 
    & Tent   & 81.19 & 77.95  & 75.43 & 68.13 & 55.15 & 56.09 \\
    & CoTTA  & 81.61 & 74.82 & 70.38 & 65.94 & 50.67 & 51.24 \\
\midrule
\multirow{2}{*}{FedProx} 
    & Tent   & 81.20 & 78.17 & 75.62 & 68.13 & 55.22  & 56.11 \\
    & CoTTA  & 81.59 & 75.03 & 70.55 & 65.95 & 50.79 & 51.28 \\
\midrule
\multirow{2}{*}{pfedGraph} 
    & Tent   & 81.19 & 77.85 & 75.49 & 68.11 & 55.16 & 55.98 \\
    & CoTTA  & 81.64 & 74.71 & 70.45 & 65.99 & 50.71 & 51.12 \\
\midrule
\multirow{2}{*}{FedAmp} 
    & Tent   & 81.19 & 78.02  & 75.53 & 68.15 & 55.39  & 56.13 \\
    & CoTTA  & 81.61 & 74.95 & 70.51 & 65.95 & 50.94 & 51.31 \\
\bottomrule
\end{tabular}
}
\end{table*}

\paragraph{Notch High Entropy (NHE) Attack Objective \cite{su2024adversarial}:}

Self-training TTA methods such as TENT and RPL are highly vulnerable to high-entropy OOD inputs, motivating entropy-maximizing attacks such as TePA. However, maximizing entropy alone does not guarantee incorrect pseudo-labels, as the model may still place probability on the true class. To overcome this, we introduce the \emph{Notch High Entropy (NHE) Attack}, which constructs a notched target distribution $Q$ by setting $Q_y = 0$ and uniformly distributing mass across all incorrect classes. The corresponding loss is
\begin{equation}
\mathcal{L}_{\mathrm{NHE}}(\mathcal{D}^{p}_{a,t}; \theta)
= \frac{1}{|\mathcal{D}^{p}_{a,t}|}
\sum_{x \in \mathcal{D}^{p}_{a,t}}
\mathrm{CE}\!\left(h(x;\theta),\, Q(x)\right),
\end{equation}
which enforces high entropy while forcing the model toward systematically wrong predictions, thereby amplifying degradation during entropy-based TTA.

\paragraph{Balanced Low-Entropy (BLE) objective \cite{su2024adversarial}:} To attack entropy-minimization TTA, we construct class-balanced low-entropy targets that nudge the model toward confident but incorrect predictions across classes (avoiding class-collapse that is easily detected). Let $q(x)$ be a target pseudo-label distribution for $x$ chosen such that the true class has near-zero mass and the remaining mass is distributed to non-true classes in a class-balanced fashion. The BLE loss on $\mathcal{B}$ under model $\theta$ is
% \begin{equation}
%     \mathcal{L}_{\mathrm{BLE}}(\mathcal{D}_{a,t}^p;\theta) \;=\; \frac{1}{|\mathcal{D}_{a,t}^p|}\sum_{x\in\mathcal{D}_{a,t}^p} \mathrm{CE}\big( h(x;\theta), q(x) \big) + \gamma \, \mathcal{R}_{\mathrm{conf}}(h(\cdot;\theta)),
%     \label{eq:ble_obj}
% \end{equation}

\begin{equation}
\begin{split}
    \mathcal{L}_{\mathrm{BLE}}(\mathcal{D}_{a,t}^p;\theta) = \frac{1}{|\mathcal{D}_{a,t}^p|}\sum_{x\in\mathcal{D}_{a,t}^p} &\mathrm{CE}\big( h(x;\theta), q(x) \big) \\
    &+ \gamma \, \mathcal{R}_{\mathrm{conf}}(h(\cdot;\theta)),
\end{split}
    \label{eq:ble_obj}
\end{equation}
where $\mathrm{CE}$ is cross-entropy and $\mathcal{R}_{\mathrm{conf}}$ is a regularizer that encourages global class balance of the induced confident labels (e.g., negative entropy of the empirical class-frequency vector) with weight $\gamma$. Minimizing~\eqref{eq:ble_obj} makes the model produce low-entropy but systematically wrong predictions after adaptation, which harms clients that use entropy minimization or pseudo-labeling.

% \subsection{Combined Optimization Objective}
% The attacker optimizes poisoned inputs (pixel perturbations or small input edits) to maximize a surrogate outer loss computed on $\hat{\theta}^{t+1}(\Delta\theta_a)$ while penalizing feature mismatch and enforcing stealth constraints. Concretely, letting $\mathcal{L}_{\mathrm{surr}}(\hat{\theta})$ denote the surrogate outer objective (instantiated as either $

\section{Result \& Discussion}

\subsection{Experimental Setup}
\paragraph{Datasets:} We conduct the experiments on two corruption benchmark datasets: CIFAR-10-C and CIFAR-100-C. These datasets were created by applying 15 different image corruptions (e.g., Gaussian noise, blur, brightness) at five discrete severity levels from 1 to 5 on the test sets of the CIFAR-10 and CIFAR-100 datasets, respectively. The corruptions at different severity levels simulate domain shifts in the data, which is a major challenge in the real world. In this work, we focus exclusively on the worst-case noise
level: severity-5, thus simulating maximal domain shift to stress-test the model’s robustness.

%%%% BOTH TENT & COTTA %%%%%
\paragraph{Implementation details:} We utilize the WideResNet-28 and ResNeXt-29 models for CIFAR-10-C and CIFAR-100-C datasets, respectively, under 15 different corruptions at severity level 5. We test the robustness of two popular Test-Time Adaptation (TTA) methods: TENT \cite{tent2021} and COTTA \cite{wang2022continual} under four different federated aggregation strategies: FedAvg \cite{mcmahan2017communication}, Fedprox \cite{li2020federated}, pFedGraph \cite{ye2023personalized}, and FedAMP \cite{huang2021personalized}. The batch size for each client is chosen to be 100. We perform the experiments for a total of N = 10 clients, of which M = 5 are adversarial, while the remaining 5 are benign. We introduce the term, poison ratio, $\alpha$, to indicate the ratio of samples within the adversarial clients' batch which are perturbed. For our experiments poison ratio ($\alpha$) is considered to be 0.5 for the adversarial clients, which implies that 50\% of the data in their batch is perturbed to degrade the adaptation performance of the benign clients.

\subsection{Results}
We evaluate our framework, FedPoisonTTP, across CIFAR-10-C and CIFAR-100-C benchmarks under diverse configurations of federated and test-time adaptation (TTA) setups to analyze the impact of adversarial clients on collaborative personalization and robustness. We report the results of our experiments in Table \ref{tab:federated_tta}. Here, `Source' indicates no Test-Time adaptation, where the evaluation scores of the frozen models on the test-sets are reported. While `None' indicates no federated aggregation, where the average accuracy values for each participating client (benign samples) are reported for various TTA techniques. 

\begin{figure}[ht]
    \centering

    % First row
    \begin{minipage}[t]{0.49\linewidth}
        \centering
        \includegraphics[width=\linewidth]{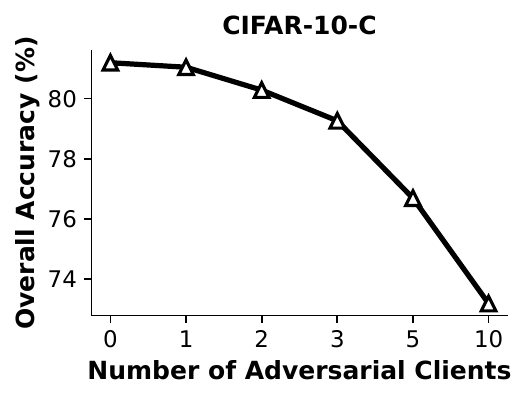}
    \end{minipage}%
    \hfill
    \begin{minipage}[t]{0.49\linewidth}
        \centering
        \includegraphics[width=\linewidth]{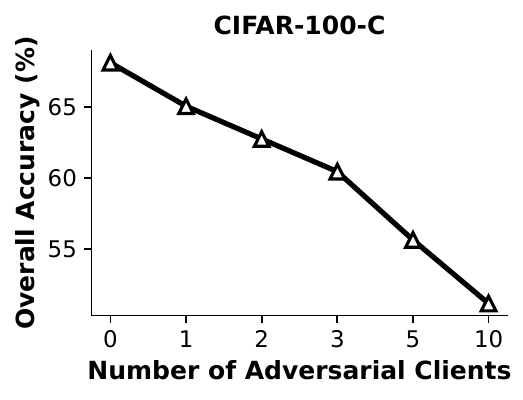}
    \end{minipage}
    \\[1ex]

    % Second row
    \begin{minipage}[t]{0.49\linewidth}
        \centering
        \includegraphics[width=\linewidth]{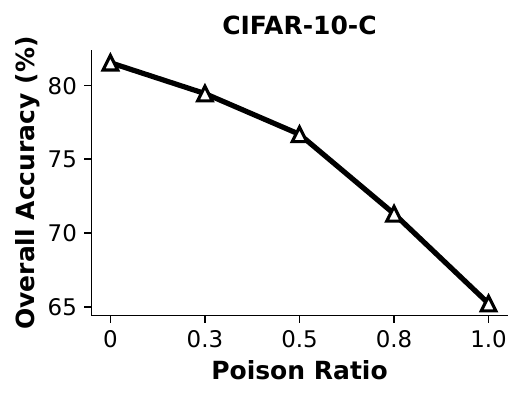}
    \end{minipage}%
    \hfill
    \begin{minipage}[t]{0.49\linewidth}
        \centering
        \includegraphics[width=\linewidth]{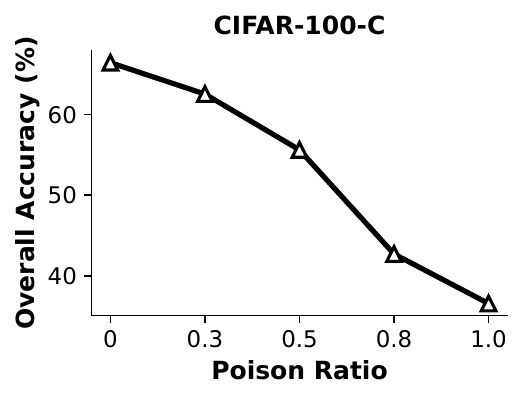}
    \end{minipage}

    % third row 
    \begin{minipage}[t]{0.49\linewidth}
        \centering
        \includegraphics[width=\linewidth]{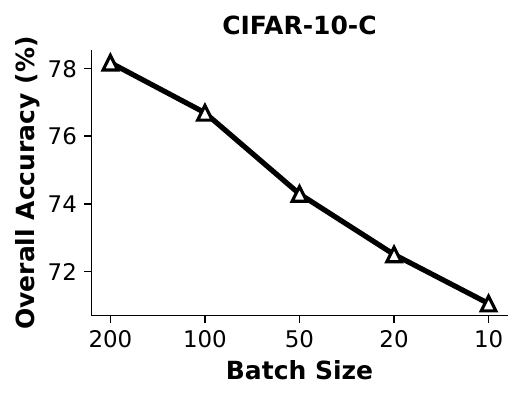}
    \end{minipage}%
    \hfill
    \begin{minipage}[t]{0.49\linewidth}
        \centering
        \includegraphics[width=\linewidth]{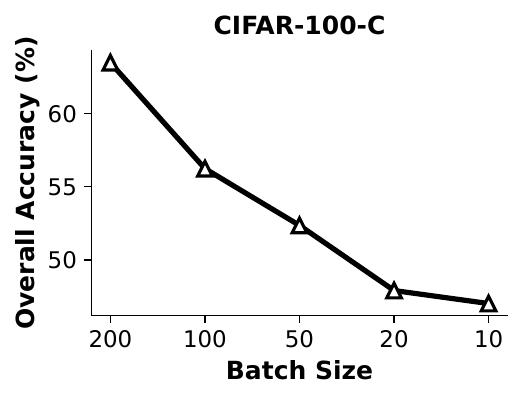}
    \end{minipage}

    \caption{Impact on overall accuracy for FedAvg-TENT setup due to varying the number of adversarial clients, ratio of malicious samples ($\alpha$), and the batch size for the participating clients. Overall accuracy reflects the average performance across the test sets of benign clients as well as the benign portion of the adversarial clients’ test data.}
    \label{fig:loss-accuracy-comparison}
    \vspace{-10pt}
\end{figure}

\begin{table*}[htbp]
\centering
\caption{Performance comparison under different poisoning objectives on CIFAR-10-C and CIFAR-100-C corruption benchmarks for FedAvg-TENT setup. Under `Benign' and `Adversarial', we report the accuracies of the benign clients and the benign data partition of the adversarial clients. `Overall' is the average accuracy of the benign clients and adversarial client's benign partition.}
\label{tab:corruption_results}
\resizebox{\textwidth}{!}{%
\begin{tabular}{l|ccc|ccc|ccc|ccc}
\toprule
\multirow{3}{*}{\textbf{Poisoning Objective}} & \multicolumn{6}{c|}{\textbf{CIFAR-10-C}} & \multicolumn{6}{c}{\textbf{CIFAR-100-C}} \\
\cmidrule(lr){2-7} \cmidrule(lr){8-13}
& \multicolumn{3}{c|}{\textbf{White box Attack}} & \multicolumn{3}{c|}{\textbf{Grey box Attack}} & \multicolumn{3}{c|}{\textbf{White box Attack}} & \multicolumn{3}{c}{\textbf{Grey box Attack}} \\
\cmidrule(lr){2-4} \cmidrule(lr){5-7} \cmidrule(lr){8-10} \cmidrule(lr){11-13}
& Overall & Benign & Adversarial & Overall & Benign & Adversarial & Overall & Benign & Adversarial & Overall & Benign & Adversarial \\
\midrule
NHE & 73.47 & 77.44 & 69.50 & 74.09 & 77.47 & 70.71 & 55.57 & 58.60 & 52.54 & 55.24 & 58.30 & 52.18 \\
NHE + Distribution Regularization & 76.25 & 77.41 & 75.10 & 76.69 & 77.95 & 75.43 & 55.42 & 54.58 & 56.26 & 55.62 & 55.15 & 56.09 \\
BLE & 74.20 & 80.81 & 67.58 & 74.54 & 80.84 & 68.23 & 62.48 & 66.63 & 58.33 & 63.04 & 66.80 & 59.29 \\
BLE + Distribution Regularization & 78.14 & 80.19 & 68.23 & 78.55 & 80.40 & 76.70 & 65.68 & 66.92 & 64.30 & 66.10 & 67.29 & 64.92 \\
MaxCE & 77.62 & 80.97 & 74.26 & 77.54 & 81.02 & 74.06 & 63.36 & 66.58 & 60.15 & 63.49 & 66.64 & 60.35 \\
Tepa & 77.68 & 80.85 & 74.51 & 75.24 & 78.99 & 71.49 & 54.07 & 56.35 & 51.78 & 54.32 & 56.86 & 51.77 \\
DIA & 78.56 & 81.06 & 76.07 & 78.28 & 81.03 & 75.52 & 62.75 & 65.91 & 59.60 & 62.59 & 65.91 & 59.28 \\
Distribution Regularization & 79.62 & 81.31 & 77.94 & 79.66 & 81.38 & 77.95 & 64.21 & 65.05 & 63.37 & 64.25 & 65.08 & 63.43 \\
\bottomrule
\end{tabular}%
}
\end{table*}

\begin{figure}[t]
    \centering
    \includegraphics[width=0.95\columnwidth]{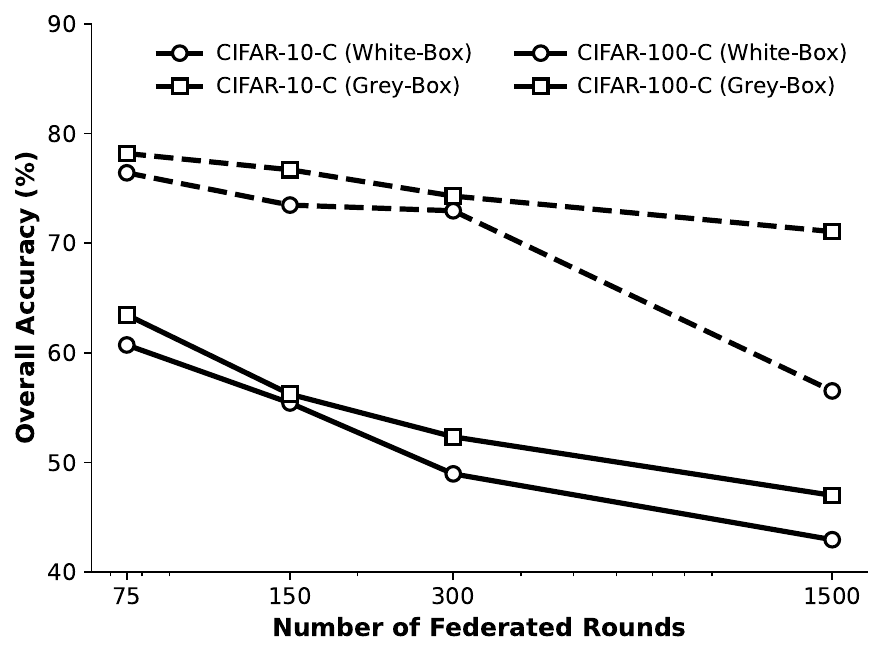}
    \caption{Overall Accuracy (\%) vs. Number of Federated Rounds.
    Performance on the benign clients and the benign partition of the adversarial clients (overall accuracy) on CIFAR-10-C (dashed lines) and CIFAR-100-C (solid lines) under White-Box (black) and Grey-Box (orange) attacks.}
    \label{fig:accuracy_vs_rounds}
\end{figure}

\paragraph{Degradation of Federated TTA under Poisoning Attacks.}
Table \ref{tab:federated_tta} reports the performance of four federated learning methods: FedAvg, FedProx, pFedGraph, and FedAMP, combined with the aforementioned test-time adaptation schemes: TENT and COTTA. Here, `clean' indicates the case of no adversarial clients, i.e., all clients are benign. For the attack scenario, we show the accuracies for the benign clients and the benign partition (50\% data in the batch) in the test set for the adversarial clients. As expected, in clean conditions, both TENT and COTTA yield strong accuracies (around 81\% on the CIFAR-10-C and around 68\% on CIFAR-100-C) for all the Fed-TTA setups. When five out of ten clients become adversarial and poison half (poison ratio, $\alpha$ = 0.5) of their local batches, the test-time accuracy across all settings is significantly reduced. For instance, adaptation via TENT under FedAvg aggregation has a performance drop from 81.19\% to 77.95\% on CIFAR-10-C and 68.13\% to 55.15\% on CIFAR-100-C, respectively, for the benign clients. The federated adaptations under COTTA show significant degradation in performance due to the participation of adversarial clients, as this strategy updates a higher number of parameters compared to TENT. For COTTA, we see the performance drop from 81.61\% to 74.82\% on CIFAR-10-C, which is a decline of 6.79\% accuracy. While keeping the same configuration, for CIFAR-100-C we observe that the presence of adversarial clients is even more detrimental, with performance dropping from 65.14\% to 50.67\%, resulting in a notable performance gap of 14.47\% for the benign clients.

\paragraph{Impact Across Datasets.}
FedPoisonTTP demonstrates performance degradation for both CIFAR-10-C \& CIFAR-100-C datasets. However, the performance drop is more prominent on CIFAR-100-C due to its higher class granularity and sensitivity to domain shifts. On average, FedPoisonTTP decreases benign clients’ accuracy by 3.03–3.34\% on CIFAR-10-C, compared to 12.76–12.98\% under TENT adaptation. The pronounced degradation on CIFAR-100-C underscores the practical impact of FedPoisonTTP, as the dataset’s larger number of classes and finer-grained semantic structure better reflect real-world scenarios where models must generalize across diverse domains.

\subsection{Ablation Studies}

We further analyze the effect of varying the number of adversarial clients, poison ratio ($\alpha$), and batch size on the overall performance of the Test-time setup. The overall accuracy, in this case, is the average accuracy of the test set of the benign clients and the benign partition of the adversarial clients. Figure \ref{fig:loss-accuracy-comparison} shows the impact of varying the number of adversarial clients, poison ratio, and batch size on the overall accuracy for the FedAvg-TENT setup.

\paragraph{Analyzing the effects of various poisoning objectives}
In Table \ref{tab:corruption_results}, we show the impact of several poisoning objectives under both white-box and grey-box (surrogate model) settings on CIFAR-10-C and CIFAR-100-C. We consider the FedAvg-TENT setup, keep the number of malicious clients equal to 5, poison ratio equal to 0.5, and a batch size of 100 for each client. A consistent trend is visible across all objectives: white-box attacks produce stronger degradation than grey-box attacks because direct access to model parameters enables more precise gradient-based manipulation. However, the relative ranking of objectives remains largely similar across both settings, indicating stable transferability of the attack behavior even under surrogate-model constraints.

Among the objectives, Distribution Regularization combined with NHE/BLE yields some of the strongest overall degradation patterns while maintaining minimal discrepancy between benign and adversarial partitions. This aligns with the intended effect of keeping poisoned samples statistically aligned with benign ones, allowing their influence to propagate more effectively through TTA updates, while diminishing the chances of attack discovery.

For both CIFAR-10-C and CIFAR-100-C, objectives such as DIA, MaxCE, and Tepa show substantial drops in adversarial-partition accuracy, while maintaining relatively higher benign accuracy. Therefore, the effect of the poisoned samples propagates less for the benign clients using these objectives. The NHE objective, on the other hand, shows a significant performance drop for the benign clients. On the more challenging CIFAR-100-C benchmark, all attacks are significantly more noticeable, with lower overall accuracy values across every objective and attack setting. This reflects the increased sensitivity of the higher-class dataset, where the model adaptation process is more easily perturbed. Finally, we see that the NHE objective, combined with distribution regularization, consistently shows maximum performance degradation among all the objectives, for both datasets under all the settings. Furthermore it is also noticeable that combining distribution regularization with the main objective improves the impact of the attack.

\paragraph{Effect of Adversarial Participation.}
Figure \ref{fig:loss-accuracy-comparison} (top) shows how the overall accuracy varies with the increasing number of adversarial clients for a fixed poison ratio of $\alpha$=0.5. The performance on CIFAR-10-C falls from 81.19\% to 73.19\% as the number of adversarial clients increases. As discussed earlier, due to having a higher number of classes, this impact is more noticeable on the CIFAR-100-C dataset where the accuracy falls from 68.13\% to 51.13\%, yielding a performance decline of 17\%. This demonstrates that even a single compromised client can cause a noticeable drop in performance, underscoring the susceptibility of Fed-TTA to localized poisoning attacks.
% \vspace{-1.5em}  % Adjust space as needed
\paragraph{Effect of Poison Ratio.} Figure \ref{fig:loss-accuracy-comparison} (middle) depicts the implications of increasing the poison ratio ($\alpha$), keeping the number of malicious clients (M) fixed to 5. We observe a drop of around 17.34\% and 29.86\% accuracy values, compared to the clean scenario, on the CIFAR-10-C and CIFAR-100-C benchmarks, respectively. This implies that increasing the amount of perturbed samples across the malicious clients is significantly detrimental to the collaborative adaptation scheme. Therefore, compared to increasing the number of adversarial clients, increasing the number of poisoned samples can significantly deteriorate the performance of the benign clients.
\paragraph{Effect of batch size}
Figure \ref{fig:loss-accuracy-comparison} (bottom) reports the impact of varying the batch size for all clients while keeping the number of clients fixed (10 in total; 5 of them are adversarial) and using a poison ratio of 0.5. We vary the per-client batch size from 200 down to 10 and evaluate the federated adaptation performance on the CIFAR-10-C and CIFAR-100-C datasets. As the batch size decreases, the accuracy shows a declining trend for both datasets. This trend arises because smaller batches render each client’s update more sensitive to the poisoned samples injected by the adversarial participants. When the batch size is large, the poisoned portion (50\% of the batch for malicious clients) is diluted by a greater number of overall samples, causing the adversarial influence to be significantly reduced. For instance, on both CIFAR-10-C \& CIFAR-100-C datasets we see high accuracy values of 78.17\% and 63.47\%, respectively, for a batch size of 200. Conversely, for a batch size of 10, the performances on both CIFAR-10-C \& CIFAR-100-C drop to 71.05\% and 47.01\% values, respectively, with the adversarial effect being more prominent on CIFAR-100-C. Therefore, with small batches, each poisoned example contributes a larger proportion of the adaptation signal, amplifying the disruption caused during test-time updates.

\paragraph{Analysing impact of attacks over various Federated Rounds}
In Fig. \ref{fig:accuracy_vs_rounds}, we report the overall classification accuracy with respect to the varying number of federated rounds. The batch size per client is varied over the values: 10, 50, 100, and 200, which correspond to the number of federated rounds: 1500, 300, 150, and 75. A reduction in the local batch size increases the total number of federated rounds, thereby providing additional opportunities for the adversarial updates to propagate through repeated model aggregation. This effect appears consistently across both datasets. Due to having a higher number of classes, this effect is more prominent for the CIFAR-100-C dataset.

\section{Conclusion}
This work introduces FedPoisonTTP, a realistic grey-box poisoning framework designed to examine the vulnerability of federated test-time personalization. By modeling an adversarial client with limited visibility and constrained poisoning capabilities, we highlight how test-time adaptation, particularly entropy-based methods, can be exploited to propagate harmful updates across collaborative learning rounds. Our empirical benchmark, spanning multiple federated aggregation methods combined with online test-time adaptation, demonstrates that even constrained adversarial participants can induce consistent and significant reductions in post-adaptation accuracy. Through extensive ablations, we further isolate the key factors that influence the severity of the attack, including the number of adversarial clients, poisoning ratio, and batch size. Overall, our findings underscore a fundamental and concerning security gap in federated test-time adaptation, while also providing a solid foundation and valuable insights for future research aimed at developing principled defenses, enhanced robustness, and more resilient training procedures.

{
    \small
    \bibliographystyle{ieeenat_fullname}
    \bibliography{main}
}

\clearpage
\appendix

\section{White-Box and Grey-Box Attack Settings}

\subsection{White-Box Attacks}
In a white-box scenario \cite{szegedy2013intriguing}, \cite{tramer2017ensemble}, \cite{wu2023uncovering},  the adversary possesses full access to the online model parameters $\theta_t$, the benign users' test samples $B_b$, and the gradients used during adaptation. This allows direct optimization of poisoned inputs against the actual online TTA dynamics:
\begin{equation}\label{eq:whitebox}
\min_{B_a} 
\ \mathbb{E}_{(x,y)\in B_b}
\left[
L_{\text{atk}}\!\left(h(x;\theta_t^{\ast}(B_a \cup B_b)),y\right)
\right],
\end{equation}
where $\theta_t^{\ast}$ denotes the updated model after TTA on both poisoned and benign samples. Although analytically convenient, this setting is unrealistic in decentralized or federated deployments where neither benign samples nor online parameters are exposed to adversaries.

\subsection{Grey-Box Attacks}
In a grey-box setting \cite{chen2017zoo}, \cite{ilyas2018black}, \cite{ru2019bayesopt}, adversaries cannot access benign users' samples or observe online model parameters $\theta_t$. Instead, they rely on a surrogate (distilled) model $\hat{\theta}_t$ that approximates the online model's behavior. The attack objective becomes:
\begin{equation}\label{eq:greybox}
\begin{aligned}
\min_{B_a} \quad 
& \frac{1}{|B_{ab}|}
\sum_{x_i \in B_{ab}}
L_{\text{atk}}\!\left(
x_i;\hat{\theta}'_t(B_t)
\right), \\[6pt]
\text{s.t.} \quad 
& B_t = B_a \cup B_{ab}.
\end{aligned}
\end{equation}
where $B_{ab}$ denotes the adversary’s own clean samples. Forwarding $B_t$ jointly can degrade the adaptation process. Here, feature-level distribution consistency is also enforced so that poisoned samples remain statistically aligned with benign ones during adaptation.

For comparison, Fig. \ref{fig:accuracy_vs_rounds} includes both white-box and grey-box (surrogate-model) attacks. As expected, the white-box variant exhibits stronger degradation because the attacker has full access to the model parameters, which is not a realistic assumption but serves as an upper bound for evaluation. The grey-box attack remains effective under limited knowledge, demonstrating its transferability. In both settings, the degradation is more pronounced for CIFAR-100-C, reflecting the greater difficulty and sensitivity of the higher-class dataset to test-time poisoning.

This experiment demonstrates that increasing the frequency of global aggregation (i.e., running more federated rounds) amplifies the influence of poisoned test-time updates and strengthens both attack variants.

\section{Various Attack Objectives}
Several attack objectives can be used to generate adversarial samples during test-time adaptation. These objectives guide gradient-based procedures such as PGD to create high-entropy disturbances, low-entropy misclassifications, or feature-consistent perturbations that influence the model’s adaptation behavior.

\subsection{Notch High-Entropy (NHE) Objective}
NHE constructs a target distribution $Q$ that assigns zero probability to the correct class and distributes uniform probability across all other classes. Minimizing the cross-entropy with respect to this target forces predictions away from the ground-truth and increases output entropy:
\begin{equation}\label{eq:nhe}
\begin{aligned}
L^{\text{NHE}}_{\text{atk}}(\tilde{x}_i)
&=
-\sum_{k} Q_{ik} \log h_k(\tilde{x}_i), \\[6pt]
\text{where} \quad
Q_{ik}
&=
\begin{cases}
0, & k = y_i, \\[4pt]
\frac{1}{K-1}, & k \neq y_i.
\end{cases}
\end{aligned}
\end{equation}

This design produces high-entropy, label-divergent predictions that strongly influence entropy-based TTA updates.

\subsection{Balanced Low-Entropy (BLE) Objective}
BLE addresses the class-collapse problem common in low-entropy attacks. It uses a moving-average confusion matrix $C$ and a label-mapping matrix $M \in \{0,1\}^{K\times K}$ that assigns each class a distinct target class. The objective encourages confident but class-balanced incorrect predictions:
\begin{equation}\label{eq:ble}
L^{\text{BLE}}_{\text{atk}}(\tilde{x}_i)
=
-
\sum_{k}
\mathbf{1}\!\left(
k = \arg\max_{q \neq y_i} M_{y_i,q}
\right)
\log h_k(\tilde{x}_i).
\end{equation}
By enforcing balanced misclassification, BLE avoids collapse into a single dominant wrong class and maintains stable target selection across all classes.

% \subsection{MaxCE Objective}
% MaxCE maximizes the standard cross-entropy loss applied to the true label:
% \begin{equation}\label{eq:maxce}
% L^{\text{MaxCE}}(\tilde{x}_i) = -\log h_{y_i}(\tilde{x}_i).
% \end{equation}
% This objective decreases the model’s confidence in the correct class and produces strong, direct gradients that guide the construction of adversarial perturbations.

\subsection{MaxCE Objective}
MaxCE objective \cite{madry2017towards} maximizes the standard cross-entropy loss with respect to the true label:
\begin{equation}\label{eq:maxce}
L^{\text{MaxCE}}(\tilde{x}_i) = -\log h_{y_i}(\tilde{x}_i).
\end{equation}
This objective drives the model’s prediction to reduce confidence in the correct class by pushing up the loss on that class. In practice, a projected gradient method (e.g., PGD) is used to find perturbations $\delta$ within a constrained norm ball that maximize this cross-entropy loss, matching the attack strategy in \cite{madry2017towards} for adversarial robustness.

\subsection{TePA Entropy-Maximization Objective}
TePA \cite{cong2024test} generates adversarial samples by maximizing prediction entropy:
\begin{equation}\label{eq:tepa}
L^{\text{TePA}}(\tilde{x}_i)
=
-\sum_{k} h_k(\tilde{x}_i)\log h_k(\tilde{x}_i).
\end{equation}
This objective encourages uniform output distributions, which destabilize entropy-based or confidence-based TTA updates by reducing prediction certainty.

\subsection{Distribution Regularization (Feature Consistency)}
Feature-level distribution regularization aligns the feature statistics of poisoned data with those of benign data. Gaussian distributions are fitted at each selected feature layer using mean–covariance pairs $(\mu_i^l,\Sigma_i^l)$ for benign samples and $(\tilde{\mu}_i^l,\tilde{\Sigma}_i^l)$ for poisoned samples:
\begin{equation}\label{eq:lreg}
L_{\text{reg}}
=
\frac{1}{L}
\sum_{l=1}^{L}
\mathrm{KLD}\!\left(
\mathcal{N}(\mu_i^l,\Sigma_i^l)
\parallel
\mathcal{N}(\tilde{\mu}_i^l,\tilde{\Sigma}_i^l)
\right).
\end{equation}
This alignment ensures that poisoned samples remain in-distribution, allowing their influence to transfer to unseen benign samples during adaptation.

\subsection{DIA (Distribution Invading Attack)}
In DIA \cite{wu2023uncovering}, poisoning is formulated as a bilevel optimization problem in which poisoned samples are crafted (outer problem) so that, after test-time adaptation (inner problem) on the mixed batch, the model incurs large loss on the benign samples. The bilevel formulation is:
\begin{equation}\label{eq:dia_bilevel}
\begin{aligned}
\min_{B_a}\quad & \mathbb{E}_{(x,y)\in B_b}\big[ L_{\text{atk}}\big(h(x;\theta^\ast(B_a\cup B_b)), y\big)\big],\\[4pt]
\text{s.t.}\quad & \theta^\ast(B_a\cup B_b) = \arg\min_{\theta} L_{\text{TTA}}\big(B_a\cup B_b;\theta\big),
\end{aligned}
\end{equation}
where $B_a$ denotes the poisoned subset and $B_b$ the benign subset in the same test batch. In practice the inner adaptation is approximated and the bilevel problem is converted into a single-level objective that is solved by projected gradient methods; a common approximation replaces $\theta^\ast$ with the current model parameters (or a surrogate) and optimizes:
\begin{equation}\label{eq:dia_single}
\min_{B_a}\; \frac{1}{|B_b|}\sum_{(x,y)\in B_b} L_{\text{atk}}\!\big(h(x;\theta_{\text{approx}}(B_a)), y\big),
\end{equation}
with constrained optimization (e.g., PGD) used to enforce perturbation bounds. DIA supports both targeted (flip a chosen sample to a target label) and indiscriminate (degrade overall performance on benign data) variants, and uses projected gradient steps to synthesize the malicious examples.

% WARNING: do not forget to delete the supplementary pages from your submission 
% \input{sec/X_suppl}

\end{document}